\documentclass{PoS}

\title{The EUSO@TurLab: Test of Mini-EUSO Engineering Model}

\ShortTitle{EUSO@TurLab\-MEEM}

\author{\small\speaker{H.~Miyamoto}$^{1,3}$, M.~Battisti$^{1,3}$, A.~Belov$^{4}$, M.E.~Bertaina$^{1,3}$, F.~Bisconti$^1$, R.~Bonino$^{1,3}$, S.~Blin-Bondil$^{5}$, F.~Cafagna$^{6}$, G.~Cambi\`e$^{7,8}$, F.~Capel$^9$, R.~Caruso$^{10,11}$, M.~Casolino$^{7,8,12}$, A.~Cellino$^{1,2}$, I.~Churilo$^{13}$, G.~Contino$^{10,11}$, G.~Cotto$^{1,3}$, A.~Djakonow$^{14}$, T.~Ebisuzaki$^{12}$, F.~Fausti$^{1,3}$, F.~Fenu$^{1,3}$, C.~Fornaro$^{15}$, A.~Franceschi$^{16}$, C.~Fuglesang$^{9}$, D.~Gardiol$^2$, P.~Gorodetzky$^{17}$, F.~Kajino$^{18}$, P.~Klimov$^{4}$, L.~Marcelli$^{7}$,  W.~Marsza{\l}$^{14}$, M.~Mignone$^{1}$,  A.~Murashov$^{5}$, T.~Napolitano$^{16}$, G.~Osteria$^{19}$, M.~Panasyuk$^{4}$, E.~Parizot$^{17}$, A.~Poroshin$^{4}$, P.~Picozza$^{7,8}$, L.W.~Piotrowski$^{12}$, Z.~Plebaniak$^{14}$, G.~Pr\'ev\^ot$^{17}$, M.~Przybylak$^{14}$, E.~Reali$^{8}$, M.~Ricci$^{16}$, N.~Sakaki$^{12}$,
 K.~Shinozaki$^{1,3}$, G. Suino$^{1,3}$, J.~Szabelski$^{14}$, Y.~Takizawa$^{12}$,  
M.~Tra\"{i}che$^{20}$, and S.~Turriziani$^{12}$ for the JEM-EUSO Collaboration\\
$^1$INFN Turin, Italy; 
$^2$OATo - INAF Turin, Italy; 
$^3$University of Turin, Italy; 
$^4$SINP, Lomonosov Moscow State University, Moscow, Russia; 
$^5$Omega, Ecole Polytechnique, CNRS/IN2P3, Palaiseau, France, 
$^6$INFN Bari, Italy; 
$^7$INFN Roma Tor Vergata, Italy; 
$^8$University of Roma Tor Vergata, Italy; 
$^9$KTH Royal Institute of Technology, Stockholm, Sweden;
$^{10}$University of Catania, Italy; 
$^{11}$INFN Catania, Italy; 
$^{12}$RIKEN, Wako, Japan; 
$^{13}$Russian Space Corporation Energia, Moscow, Russia; 
$^{14}$National Centre for Nuclear Research, Lodz, Poland; 
$^{15}$UTIU Rome, Italy; 
$^{16}$INFN - Laboratori Nazionali di Frascati, Italy; 
$^{17}$APC, Univ Paris Diderot, CNRS/IN2P3, CEA/Irfu, Obs de Paris, Sorbonne Paris Cit\'e, France; $^{18}$Konan University, Kobe, Japan; 
$^{19}$INFN Naples, Italy; 
$^{20}$Centre for Development of Advanced Technologies (CDTA), Algiers, Algeria \\
}

\abstract{
The TurLab facility is a laboratory, equipped with a 5 m diameter and 1 m depth rotating tank, located in the Physics Department of the University of Turin.
Originally, it was mainly built  to study systems of different scales where rotation plays a key role in the fluid behavior such as in atmospheric and oceanic flows.
 In the past few years the TurLab facility has been used to perform experiments related to the observation of Extreme Energy Cosmic Rays (EECRs) from space using the fluorescence technique. 
For example, in the case of the JEM-EUSO mission, where the diffuse night brightness and artificial light sources can vary significantly in time and space inside the Field of View of the telescope.
The Focal Surface of Mini-EUSO Engineering Model (Mini-EUSO EM) with the level 1 (L1) and 2 (L2) trigger logics implemented in the Photo-Detector Module (PDM) has been tested at TurLab.
Tests related to the possibility of using an EUSO-like detector for other type of applications such as Space Debris (SD) monitoring and imaging detector have also been pursued.
The tests and results obtained within the EUSO@TurLab Project on these different topics are presented.
}

\FullConference{36th International Cosmic Ray Conference -ICRC2019-\\
		July 24th - August 1st, 2019\\
		Madison, WI, U.S.A.}

\begin{document}\setcounter{page}{2}
\section{Introduction}
\subsection{TurLab@Physics Department - University of Turin}
TurLab~\cite{ref:TurLab} is a laboratory for geo-fluid-dynamics studies, where rotation is a key parameter such as Coriolis force and Rossby Number.
By using inks or particles, based on the fluid-dynamics, key phenomena such as planetary atmospheric and fluid instabilities can be reproduced in the TurLab water tank.
The tank has 5 m diameter with a capability of the rotation at a speed of 3 s to 20 min per rotation.
Also, as it is located in a very dark environment, the intensity of background light can be adjusted in a controlled condition.
\subsection{The EUSO@TurLab project}
The EUSO@TurLab project~\cite{ref:EUSOatTurLab} is a series of measurement campaign, in which we have tested several kinds of prototypes and pathfinders of the fluorescence telescopes equipped with the "EUSO electronics".
As described in the following sections, those telescopes are designed to observe the Earth's atmosphere from the stratosphere or space from the orbit of the International Space Station (ISS). 
By means of the rotating tank with the capabilities mentioned above, we have been testing the EUSO electronics such as its basic performance and the first level trigger (L1) for cosmic-rays, in view of various and even changing background conditions as well as atmospheric phenomena such as meteors and lightning that EUSO telescopes will observe.
\subsection{Mini-EUSO: A pathfinder of JEM-EUSO project}
The Mini-EUSO~\cite{ref:Mini-EUSO} is one of the pathfinders of the JEM-EUSO project~\cite{ref:JEM-EUSO} which is a concept of a space-borne fluorescence telescope to be hosted on the ISS.
Mini-EUSO has been approved by both the Russian (Roscosmos) and Italian (ASI) space agencies and is set to be launched to the Zvezda module of the ISS this year (2019). 
Mini-EUSO will perform mapping the UV emissions of the night-earth with a temporal resolution of 2.5 $\mu$s and a spatial resolution of $\sim$5 km.
It will also observe a variety of atmospheric events such as transient luminous events (TLEs) and meteors, as well as searching for strange quark matter and bio luminescence.
In addition, Mini-EUSO will be used for SD detection to verify the possibility of using an EUSO-class telescope in combination with a high-energy laser for SD remediation~\cite{ref:Toshi}.
\section{Mini-EUSO trigger}
Mini-EUSO Focal Surface (FS) consists of a PDM, where 36 Hamamatsu 64ch MultiAnode PhotoMultiplier Tubes (MAPMTs) are arrayed in a 6$\times$6 matrix, resulting in a readout of 2304 pixels.
Signals are pre-amplified and converted to digital values by the SPACIROC3 ASIC~\cite{ref:SPACIROC3} for each time window of 2.5$\mu$s, which is hereafter referred to as one Gate Time Unit (GTU).
This data is then passed to the data processing unit for data handling and storage.

The L1 trigger concerns data with a time resolution of 2.5$\mu$s (= 1 GTU) and looks for signal excess on a timescale of 20 $\mu$s (=8 GTUs).
This is the typical timescale of cosmic-ray-like events.
The data integrated every 128 GTUs (320$\mu$s) is used to determine the background level and at the same time, is passed to the L2 trigger logic as 1 L2\_GTU as described in the next paragraph.
Photon counts of each pixel are integrated over 8 consecutive GTUs and compared to the background level, which is determined by the previously integrated data described above. 
When a trigger is issued, the whole FS is read out and a packet of 128 GTUs is stored, centred on the trigger time.

\begin{figure}
\centering
\includegraphics[width=\hsize,height=14.2cm]{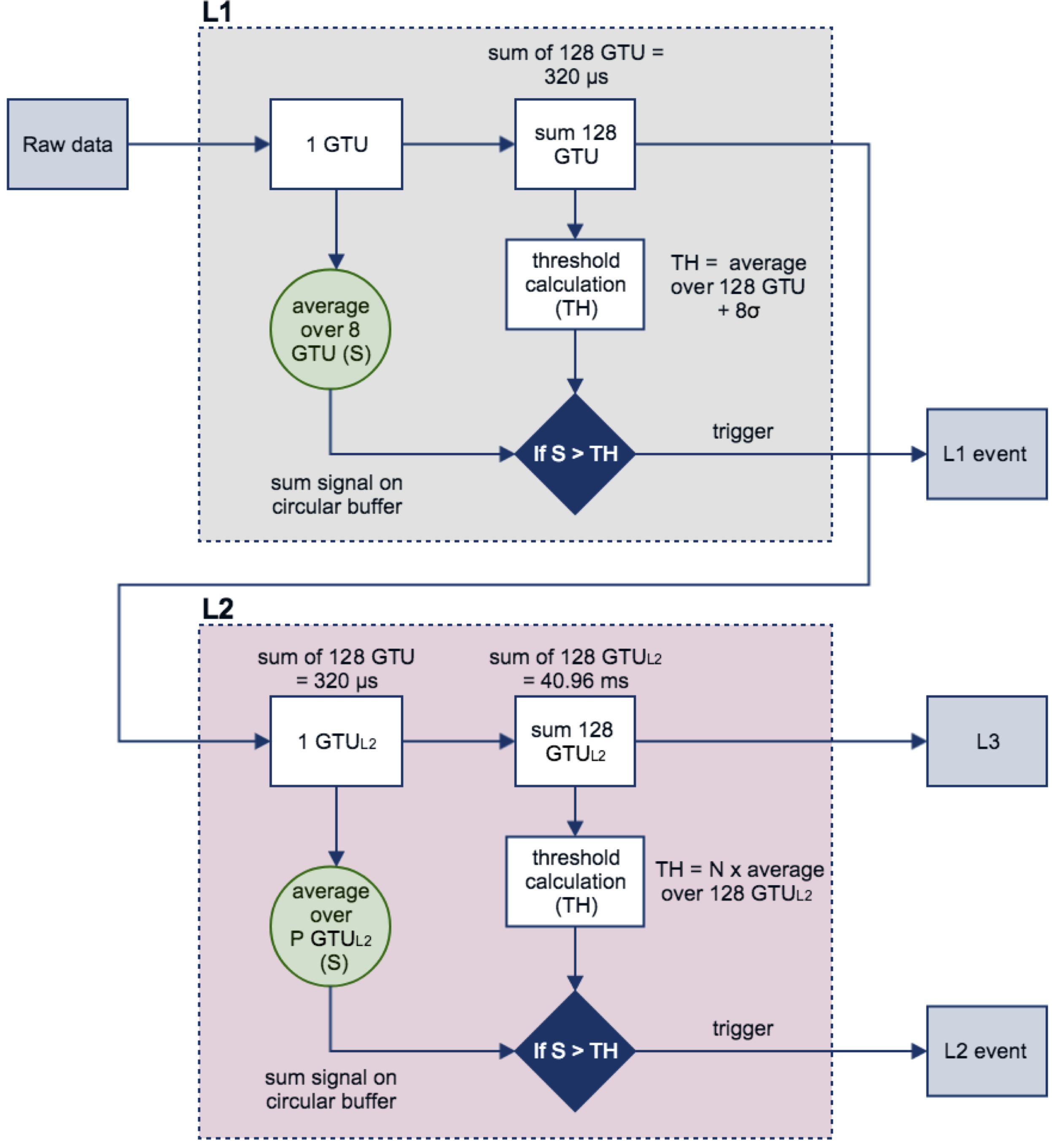}
\caption{Block diagram of L1 and L2 trigger algorithm.
\label{fig:L1L2schem}}
\end{figure}

The L2 trigger works with a similar logic as L1, but with a time resolution of 320 $\mu$s (=1 L2 GTU), well-suited to capturing fast atmospheric events such as TLEs and lightning, which have timescales of 10$\mu$s - 100 ms~\cite{ref:TLE}.
Background is set by integrating 128 L2 GTUs, which is then rescaled on 8 GTUs as shown in Fig. ~\ref{fig:L1L2schem}.
This packet of 128 L2 GTUs is then also stored as the level 3 (L3) data as 1 L3 GTU.
During the accumulation of 128 L3 GTUs (=5.24s), up to 4 events from L1, L2 are stored, then transferred to the CPU with the L3 data for formatting and storage on the disk.
In this way, a continuous and controlled readout is achieved with a resolution of 40.96 ms whilst also keeping interesting events at faster timescales.
This 40.96 ms "movie" will be used to search for meteors, SD and strange quark matter as well as for the mapping of the Earth in UV. 
\section{Mini-EUSO EM@TurLab}
In February to March 2018, the Mini-EUSO Engineering Model (Mini-EUSO EM) 
has been tested at TurLab.
After fundamental operation tests after arrival, it was hung on the ceiling above the TurLab tank to be tested for its general performance as well as for the trigger system.

\begin{figure}
\vspace{-0.2cm}
\centering
\includegraphics[width=\hsize]{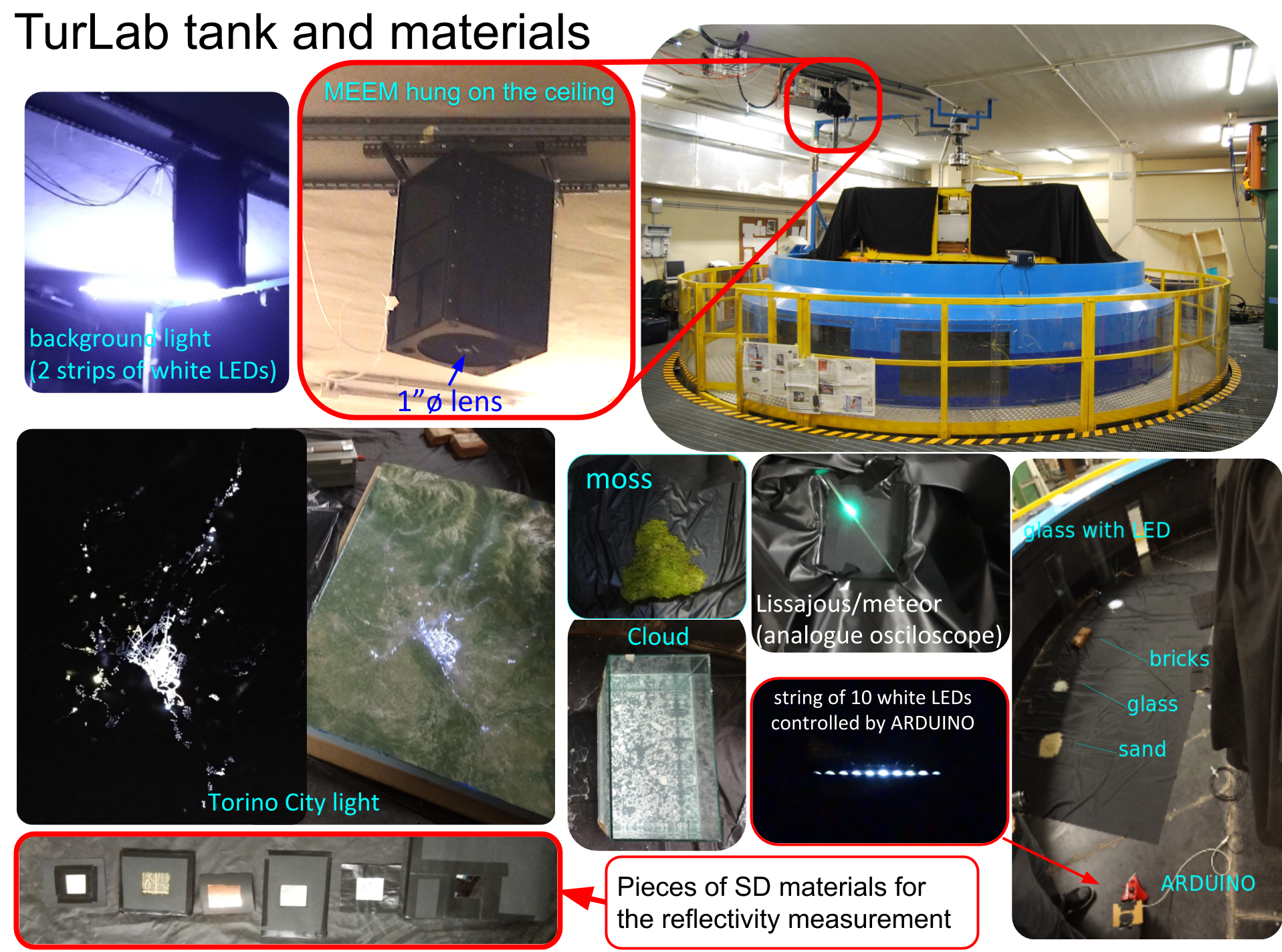}
\caption{
TurLab tank (right-top) setup with Mini-EUSO EM (top-centre) hung on the ceiling, light sources and materials to reproduce phenomena that Mini-EUSO detector will observe.
Reflections of the materias which often make Space Debris (SD) are also tested (bottom-left). 
\label{fig:TurLabMat}
}
\vspace{-0.2cm}
\end{figure}

Fig.~\ref{fig:TurLabMat} shows the setup for the TurLab measurement.
Right-top shows the TurLab tank, and on top of it, the detector to be tested is installed.
Mini-EUSO EM was set on the ceiling shown as in the area surrounded by red line and zoomed in the top-middle of the figure. A 1" $\phi$ plano-convex lens is used to focus the light this time.
All other parts of the photo show the light sources and materials reproducing the various phenomena that Mini-EUSO will observe from space, such as rocks, desert, glacier ice, cloud, forest, lightning, city light as well as meteors and cosmic-rays each of which are reproduced by means of bricks, sand, smashed glass, clusters of particles floating in the water, moss, smashed LED illuminated by blinking LED, LED light through the holes on the scaled map of Torino city, Lissajous of an analogue oscilloscope and Arduino driven LED strip.
All these materials are illuminated by diffused background light which is illuminating the ceiling above the tank to reproduce the diffused air glow in the atmosphere at the level of expected photon counts that Mini-EUSO will observe ($\sim$1 count/pix/GTU).
Rotating the tank above these light sources and materials, we reproduce the Mini-EUSO observation on the ISS orbit.
\begin{figure}
\centering
\includegraphics[width=\hsize,height=7.92cm]{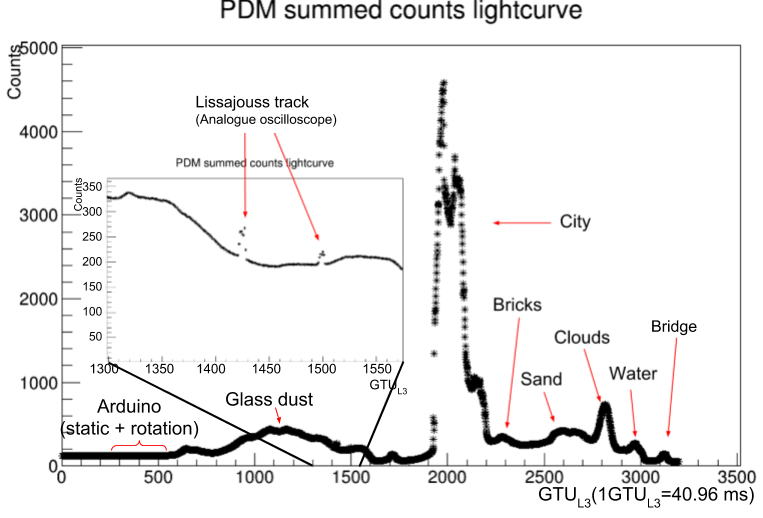}
\caption{
Light curve of L3 data for a whole tank rotation with a speed of $\sim$2 min/rot. All the materials in the FoV are recorded as a continuous "movie" in L3 data while L1 and L2 stores only the events that they are targeting to trigger in their own time resolution (up to 4 events per D3 GTU).
\label{fig:D3LC}
}
\end{figure}

The Fig. ~\ref{fig:D3LC} shows the light curve of Mini-EUSO EM L3 data for the TurLab tank rotation measurement.
Even though the data acquisition firmware at that time was not fully debugged and there were some inconvenience or was not fully functioning, we obtained many useful data for debugging and development of the later versions of the firmware.
\subsubsection*{L1 trigger}
\begin{figure*}
\centering
\includegraphics[width=0.98\hsize,height=7cm]{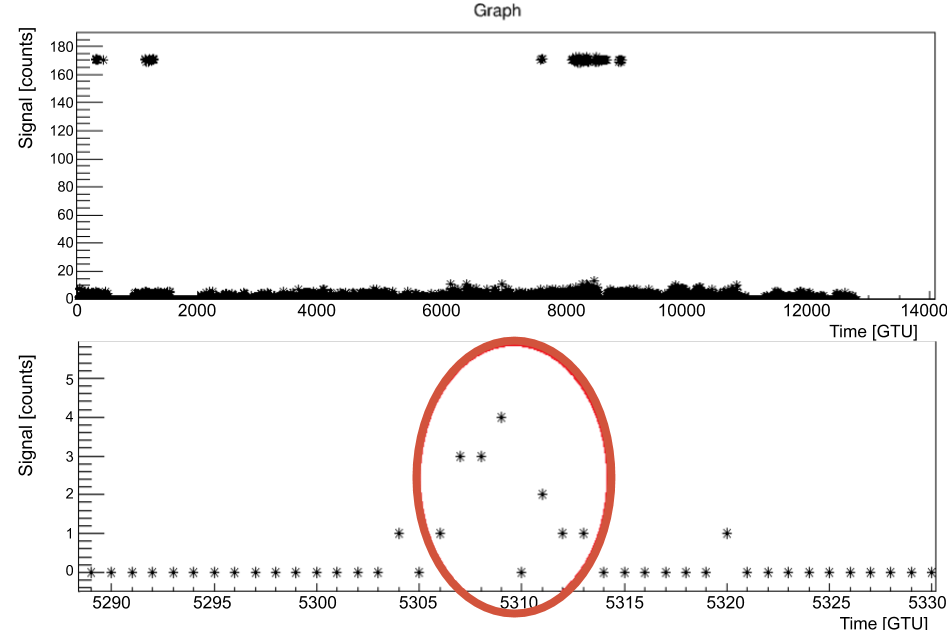}
\caption{
Top: A pixel photon counts in L1 data as a function of GTUs. Bottom: the area zoomed in around GTU 5309, where the integrated counts in 8 consecutive GTUs exceed 15 counts (in the red circled area), which is the minimum threshold when the average of the previous 320 $\mu$s is lower than 0.3 counts/GTU.
\label{fig:fakeT}
}
\end{figure*}

L1 was triggering on and storing the data of cosmic-ray-like event, which is reproduced with a strip of LEDs driven by Arduino, with an interval of 10 ms between the repetition of the events  during the tank rotation.
However, we also found that there were fake triggers with higher rate than the requirement ($\leq$0.75 Hz), roughly at every several 100 ms.
In our system, when the background is very low and if the threshold doesn't reach the level of 15 counts in 8 consecutive GTUs, L1 takes the threshold of 15 counts.
Investigating some of these unwanted trigger events, we found that there are some pixels which does not follow the Poissonian behavior due to the electric noise or due to intrinsic characteristics of the MAPMT.
For example, Fig.~\ref{fig:fakeT} shows the light curve of one of such pixels which have very low background counts.
In such a case, the minimum threshold of 15 counts is applied as the average count of the last 320 $\mu$s is very low.
In the past studies, this restriction was estimated to be sufficient against the noises in a Poissonian distribution, however, we found that there is a risk that it is not high enough against the other types of noise such as electric noise, or higher fluctuation of the counts at unstable pixels.
Fig. ~\ref{fig:fakeT} shows an example of such an event; the background counts are exceeding 15 while the average count of the last 320 $\mu$s is $\sim$0.3 counts/GTU.
To avoid this problem, we increased the threshold to 16 $\sigma$ instead of 8 $\sigma$ in the latest system as described in the following section.
\subsubsection*{L2 trigger and L3 movie data}
Unlike the L1, which still had to be improved and debugged in the logic, the L2 trigger was working as intended.
The bright event lasting relatively long such as Lissajouss track produced by an analogue oscilloscope, which gives a smilar time profile to TLEs in this case, has been triggered by L2. 
It also triggered on the city light where the light increased rapidly, but any other materials or light sources such as glass dust or Arduino did not generate a trigger. 

Also, as shown in Fig.~\ref{fig:D3LC} for the L3 data, the continuous movie was successfully stored during the measurements without any problem and each material and light source are clearly recognised. 
\subsubsection*{SD measurement at TurLab with Mini-EUSO EM}
Additional to the tests above, we also made a special setup to reproduce the SD observation at the TurLab tank with Mini-EUSO EM 
as well as the reflectivity test of the materials which usually SD consists of (shown in the left bottom of Fig. ~\ref{fig:TurLabMat}).
The details of this test will be reported in another contribution in this conference ~\cite{ref:SDmiyamoto}.
\section{Latest measurements and results}
\begin{figure}
\vspace{-0.2cm}
\centering
\includegraphics[width=0.8\hsize,height=8.5cm]{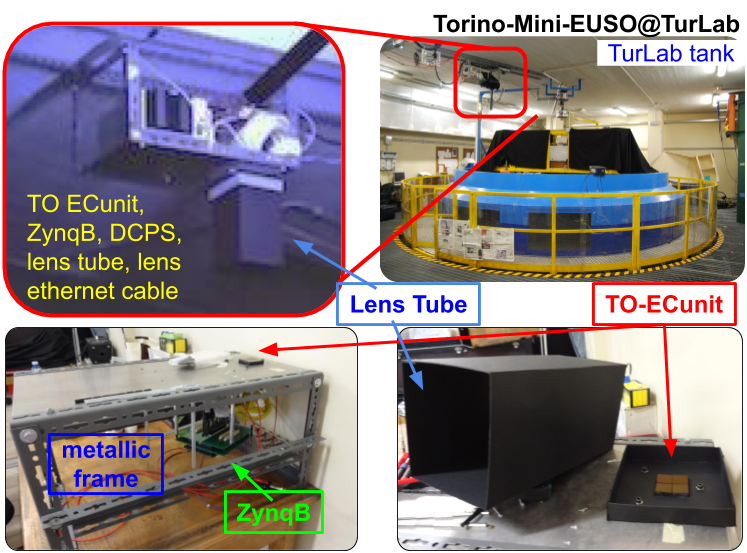}
\caption{
Torino-Mini-EUSO consists of a lens tube with a 1" $\phi$ plano-convex lens, an EC unit, front-end electronics based on SPACIROC3 ASICs, the Zinq board connected to a PC via ethernet cable, where the CPU software, a dedicated software for Mini-EUSO data processing system is installed. Electronics boards and PMTs are powered by external low and high voltage power supplies respectively.
\label{fig:TOME}
}
\vspace{-0.2cm}
\end{figure}

In the beginning of 2019, in parallel to the intensive test of Mini-EUSO EM in Rome, we build a mockup of the system in Turin. 
This "Torino MiniEUSO-like detector" consists of 4 EC units, the Zynq board which is developed for Mini-EUSO and other EUSO pathfinders, and a PC on which CPU software developed for Mini-EUSO operation has been installed (See Fig.~\ref{fig:TOME}).
The difference from the official Mini-EUSO detector as well as Mini-EUSO EM is that we do not have other subsystems such as High-Voltage (HV) boards, ancillary cameras or the Fresnel-lens based optical system.
It solely consists of the data acquisition and processing part but in an identical way to the full telescope.
It is useful not only because we can test the detector performance in parallel to the development of the Mini-EUSO detector, but also makes us easier to investigate when a problem occurs, separating from the other kind of problems caused by other subsystems.
In April to June, we performed the same measurement at TurLab with this system as we have done for Mini-EUSO EM to debug and complete the L1 algorithm.
We reconfirmed that the problem of relatively high trigger rate that we observed in the measurement of Mini-EUSO EM with the L1 trigger with 8 $\sigma$ threshold, i.e., either when we are in a very low background level, or much higher than the standard level on the contrary, the fake trigger rate increased and become higher than it is required.
Then we updated the firmware of the Zynq board, setting a threshold of 16 $\sigma$ for L1, with a value of 28 counts for the minimum threshold, unlike the case of 8 $\sigma$ (with the value of 15 counts for the minimum threshold), it triggered only the cosmic-ray-like events driven by Arduino through the whole rotation, and had almost no fake triggers.
A drawback is that it is naturally increasing the energy threshold against the cosmic-ray observation, however, still less than by a factor of 2 in this case.
Thus, we concluded that the L1 trigger with a threshold of 16 $\sigma$ should be applied for the Mini-EUSO detector operation.
\begin{figure}
\vspace{-0.2cm}
\centering
\includegraphics[width=0.9\hsize]{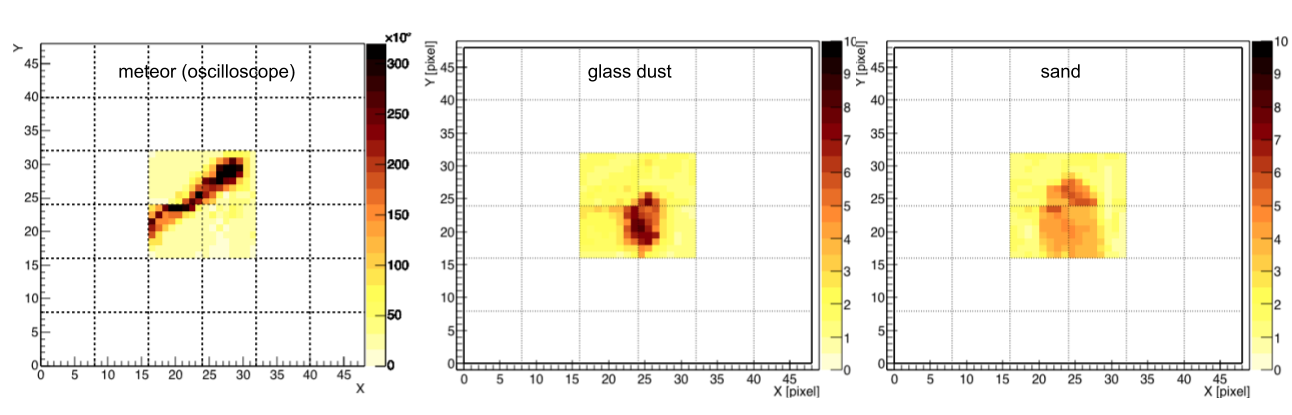}
\caption{
Examples of L3 data images of meteor, smashed glass, sand. For the meteor, it is an integrated image of 20 L3 GTUs to make the moving track visible, while the other two are those of just 1 L3 GTU. 
\label{fig:TLD3image}
}
\vspace{-0.2cm}
\end{figure}
\vspace{-0.5cm}
\section{Summary and Conclusion}
The Mini-EUSO EM has been successfully tested at TurLab in 2018. This test was useful as we could confirm that the L2 trigger algorithm and L3 data acquisition are working as expected, while we found some items to be improved and debugged for L1. With a new system, which is in principle a copy of the Mini-EUSO detector,  but only by the data acquisition and the processing part, we were able to perform further studies recently.
A new threshold of 16 $\sigma$ has been applied to the L1 trigger and now it is more robust and stronger against the non-standard (much lower or higher than $\sim$1 count/pix/GTU) background condition.
\vspace{-0.2cm}
\section*{Acknowledgements}
\vspace{-0.2cm}
\noindent 
This work was partially supported by Basic Science Interdisciplinary Research
Projects of RIKEN and JSPS KAKENHI Grant (JP17H02905, JP16H02426 and
JP16H16737), by the Italian Ministry of Foreign Affairs and International
Cooperation, by the Italian Space Agency through the ASI INFN agreement
n. 2017-8-H.0 and contract n. 2016-1-U.0, by NASA award 11-APRA-0058 in
the USA, by the
Deutsches Zentrum f\"ur Luft- und Raumfahrt, by the French space agency
CNES, the Helmholtz Alliance for Astroparticle Physics funded by the
Initiative and Networking Fund of the Helmholtz Association (Germany), by
Slovak Academy of Sciences MVTS JEMEUSO as well as VEGA grant agency project
2/0132/17,
by National Science Centre in Poland grant (2015/19/N/ST9/03708 and
2017/27/B/ST9/02162), by Mexican funding agencies PAPIIT-UNAM, CONACyT and
the
Mexican Space Agency (AEM). Russia is supported by ROSCOSMOS and the Russian
Foundation for
Basic Research Grant No 16-29-13065. Sweden is funded by the Olle Engkvist
Byggm\"astare Foundation.
\vspace{-0.2cm}
\end{document}